\documentclass{mecbic}
\providecommand{\anno}{2011} \providecommand{\firstpage}{95}
\providecommand{\lastpage}{111}
\usepackage{breakurl}        

\usepackage{graphicx}
\usepackage{latexsym}
\usepackage{amsmath}
\usepackage{graphs}

\usepackage{amssymb}
\usepackage{url}

\usepackage{multirow}
\usepackage{verbatim}

\usepackage{graphs}

\newcommand{\NonStandardSize}   {\setlength{\unitlength}{0.55cm}}

\newcommand{\stateNode}[4]{\roundnode{#1}(#2,#3)\freetext(#2,#3){$#4$}}

\def\nat{{\bf N}}

\def\mvn{{\it MV}}
\def\mvni#1{\mvn_{#1}}
\def\mvnEx1{Ex1}

\def\nodeAll{ G } 
\def\node#1{ g_{#1} } 
\def\nextSt#1{ [#1] }
\def\modelSS#1{ S_{\scriptstyle #1} }
\def\ss{Y}

\def\ssEntity#1{ \ss(#1) } 

\def\nbh{N}
\def\nbhEntity#1{\nbh(#1)} 
\def\func{F}
\def\funcEntity#1{ f_{ #1 } } 

\def\absMap{\phi}
\def\mapping#1{\absMap(#1)}

\def\updateStep{\rightarrow}

\def\PLI{{\it PL2}}
\def\PLIab{{\it APL2}}

\def\trpMV{{\it MTRP}}
\def\trpBool{{\it ATRP}}
\def\Trp{{\it Trp}}
\def\TrpExt{{\it TrpExt}}
\def\TrpE{{\it TrpE}}
\def\TrpR{{\it TrpR}}

\def\pfbox{ \hspace*{\fill} $\Box$ }

\newcommand{\CI}     {\mathit{CI}}

\newcommand{\Cro}    {\mathit{Cro}}

\def\upD#1#2{[#1]^{#2}}
\def\upStepSyn{\xrightarrow{Syn}}
\def\upStepAsy{\xrightarrow{Asy}}
\def\nxtState#1{next^{#1}}
\def\stGraphSyn#1{{\it SG}^{S}(#1)}
\def\stGraphAsy#1{{\it SG}^{A}(#1)}
\def\trSetA#1{Tr^{A}(#1)}
\def\trSetS#1{Tr^{S}(#1)}
\def\absA#1{\lhd_{A}^{#1}}
\def\setStAbs#1{\absMap^{-1}(#1)}
\def\stepTerm#1#2#3{[#1 : #2 : #3 ]}
\def\setS#1{D(#1)}

\def\stShort#1#2{st(#2,#1)}
\def\allStepTerm#1{{\it Step}(#1)}
\def\equivClass#1{[#1]^{\absMap}}
\def\clSetST#1{H(#1)}

\newcounter{defsThms}
\def\addDefn#1#2{{\refstepcounter{defsThms}\label{#2}{\bf #1  \arabic{defsThms}.~}}}

\title{Abstracting Asynchronous Multi-Valued Networks: \\
           An Initial Investigation}
\author{L. Jason Steggles
  \institute{Newcastle University, UK.}
  \email{L.J.Steggles@ncl.ac.uk}
}
\def\titlerunning{Abstracting Asynchronous Multi-Valued Networks:
     An Initial Investigation}
\def\authorrunning{L. Jason Steggles}

\begin{document}


\maketitle \pagestyle{plain} \pagenumbering{arabic}
\setcounter{page}{95}

\begin{abstract}
Multi-valued networks provide a simple yet expressive qualitative state based modelling  approach for biological systems.
In this paper we develop an abstraction theory for asynchronous
multi-valued network models that allows the state space of a model to be reduced while preserving key properties of the model.
The abstraction theory therefore provides a mechanism for coping with the state space
explosion problem and supports the analysis and comparison of multi-valued networks.
We take as our starting point the abstraction theory for synchronous multi-valued networks
which is based on the finite set of traces that represent the behaviour of such a model.
The problem with extending this approach to the asynchronous case is that
we can now have an infinite set of traces associated with a model making a simple trace inclusion
test infeasible.
To address this we develop a decision procedure for checking
asynchronous abstractions based on using the finite state graph of an
asynchronous multi-valued network to reason about its trace semantics.
We illustrate the abstraction techniques developed by considering a detailed case study based on a multi-valued network model of the regulation of tryptophan biosynthesis in {\it Escherichia coli}.
\end{abstract}


\section{Introduction}
{\it Multi-valued networks} (MVNs)
\cite{Rudell1987,Thomas1990,Thomas1995} are an expressive
qualitative modelling approach for biological systems (for example,
see \cite{Thomas1995,Chaouiya2008,Schaub2007,banksSteggles2007}).
They extend the well--known {\it Boolean network}
\cite{kauffman1969,kauffman1993} approach by allowing the state of
each regulatory entity to be within a range of discrete values
instead of just {\it true} or {\it false}. The state of each
regulatory entity is influenced by other regulatory entities in the
MVN and entities update their state using either a {\it synchronous
update strategy} \cite{kauffman1993,wuensch2002} where all entities
simultaneously update their state, or an {\it asynchronous update
strategy} \cite{Thomas1973,Harvey1997,Tournier2009} where entities
update their state independently using a non-deterministic approach.

While MVNs have shown their usefulness for modelling and understanding biological systems further
work is still needed to strengthen the techniques and tools available for MVNs.
One interesting area that needs developing is a theory for abstracting MVNs.
Abstraction techniques allow a simpler model to be identified
which can then be used to provide insight into the more complex original model.
Such techniques are well--known in the formal verification community as a means of coping
with the complexity of formal models (see for example \cite{Clarke1994,Bensalem1998,Clarke03,DSilva2008}).
The main motivation behind developing such a theory for MVNs can be summarised as follows:
\\
\\
(1) The analysis of MVNs is limited by the well--known problem
of state space explosion.
Using abstraction is one useful approach which allows analysis results from a simpler
approximate model to infer results about the original model.
\\
(2) Often several MVNs are defined at different levels of abstraction when modelling a system.
It is therefore clearly important to be able to formally relate these models using an
appropriate theory.
\\
(3) An abstraction theory would provide a basis for the step--wise refinement of MVNs.
\\
(4) Identifying an abstraction for a complex MVN provides a means of better
visualising and understanding the behaviour an MVN, giving greater insight into the system being modelled.
\\

The abstraction theory we present for asynchronous MVNs is based on
extending the synchronous abstraction theory presented in
\cite{banksSteggles2010}. We formulate a notion of what it means for
an MVN to be correctly abstracted by a simpler MVN with the same
network structure but smaller state space. The idea is to use an
abstraction mapping to relate the reduced state space of an
abstraction to the original MVN. An abstraction is then said to be
correct if its set of traces is within the abstracted traces of the
original MVN. This definition of abstraction represents an {\it
under--approximation} \cite{Clarke1994,Pelanek2006} since not all of
the behaviour of the original MVN is guaranteed to have been
captured within the abstraction. We show that this approach allows
sound analysis inferences about positive reachability properties in
the sense that any reachability result shown on an abstraction must
hold on the original model. An important result of this is that it
therefore follows that all attractors of an asynchronous abstraction
correspond to attractors in the original MVN. Note that an
alternative approach commonly used in abstraction is to use an {\it
over--approximation} \cite{Clarke1994,Pelanek2006,Clarke03} in which
false positives may occur. However, such an approach appears to be
problematic for MVNs and we discuss this further in Section
\ref{sec:asynAbs}.

The non-deterministic nature of asynchronous MVNs mean that we encounter additional
complications compared to the synchronous case;
an asynchronous MVN can have an infinite set of traces which means that
directly checking trace inclusion to check a proposed abstraction is infeasible.
We overcome these difficulties by constructing a decision procedure
for checking asynchronous abstractions that is based on the
underlying finite state graph of an MVN. We introduce the idea of
{\it step terms} which are used to denote possible ways to use sets
of concrete states to represent abstract states. The decision
procedure starts with the set of all possible step terms and then
iteratively prunes the set until either a consistent abstract
representation has been found or the set of remaining step terms is
too small to make it feasible to continue. We provide a detailed
proof that shows the decision procedure correctly identifies
asynchronous abstractions and discuss the complexity of the decision
procedure.

We illustrate the abstraction theory we develop by considering a case study
based on modelling the regulatory network that controls the biosynthesis of tryptophan by
the bacteria {\it E. coli} \cite{SenLiu1990,Santillan2001}.
Tryptophan is essential for the development of {\it E. coli} and its resource intensive synthesis is carefully controlled to ensure
its production only occurs when an external source is not available.
We investigate identifying asynchronous abstractions for an existing MVN model of this
regulatory mechanism which was developed in \cite{Simao2005}.

The paper is organized as follows. In Section \ref{sec:mvn} we
provide a brief overview of the MVN modelling framework and present
a simple illustrative example. In Section \ref{sec:asynAbs} we
formulate a notion of abstraction for asynchronous MVNs and consider
the analysis properties that can be inferred from an abstraction. In
Section \ref{sec:decProc} we present a decision procedure for
checking asynchronous abstractions and provide a detailed proof of
correctness for this procedure. In Section \ref{case} we illustrate
the theory and techniques developed by a case study based on
modelling the regulatory network that controls the biosynthesis of
tryptophan by {\it E. coli}. Finally, in Section \ref{concl} we
present some concluding remarks and discuss related work.

\section{Multi-valued Network Models}
\label{sec:mvn} In this section, we introduce {\it multi-valued
networks} (MVNs) \cite{Rudell1987,Thomas1990,Thomas1995}, a
qualitative modelling approach which extends the well-known {\it
Boolean network} \cite{kauffman1969,kauffman1993} approach by
allowing the state of each regulatory entity to be within a range of
discrete values. MVNs can therefore discriminate between the
strengths of different activated interactions, something which
Boolean networks are unable to capture. MVNs have been extensively
studied in circuit design (for example, see
\cite{Rudell1987,Mishchenko2002}) and successfully applied to
modelling biological systems (for example, see
\cite{Thomas1995,Chaouiya2008,Schaub2007,banksSteggles2007}).

An MVN consists of a set of logically linked entities $G = \{\node{1} , \ldots, \node{k} \}$ which regulate each other in a positive or negative way.
Each entity $\node{i}$ in an MVN has an associated set of discrete states $\ssEntity{\node{i}} = \{0,\dots, m_{i}\}$, for some $m_{i} \geq 1$, from which its current state is taken.
Note that a Boolean network is therefore simply an MVN in which each entity $\node{i}$ has a Boolean set of states $\ssEntity{\node{i}} = \{0, 1\}$.
Each entity $\node{i}$ also has a neighbourhood $\nbhEntity{\node{i}} = \{\node{i_{1}}, \ldots, \node{i_{l(i)}} \}$
which is the set of all entities that can directly affect its state.
A given entity $\node{i}$ may or may not
be a member of $\nbhEntity{\node{i}}$ and any entity in which
$\nbhEntity{\node{i}} = \{ \}$
is taken to be an input entity whose regulation is outside the current model.
The behaviour of each entity $\node{i}$ based on these neighbourhood interactions is
formally defined by a logical next-state function
$\funcEntity{\node{i}}$
which calculates the next-state of $\node{i}$ given the current states of the entities in its neighbourhood.

We can define an MVN more formally as follows.
\\
\\
\addDefn{Definition}{def:mvn}
An MVN $\mvn$ is a four-tuple $\mvn = (\nodeAll, \ss, \nbh, \func)$ where:
\\
i) $\nodeAll=\{ \node{1} , \dots, \node{k} \}$ is a non-empty, finite set of entities;
\\
ii) $\ss = \left( \ssEntity{ \node{1}} , \ldots, \ssEntity{\node{k}} \right)$ is a tuple of state sets,
where each $\ssEntity{\node{i}} = \{0, \ldots, m_{i}\}$, for some $m_{i} \geq 1$, is the state space for entity $\node{i}$;
\\
iii) $\nbh = \left( \nbhEntity{ \node{1}}  , \ldots, \nbhEntity{ \node{k}} \right) $ is a tuple of neighbourhoods,
such that $\nbhEntity{ \node{i}}  \subseteq \nodeAll $ is the neighbourhood of $\node{i}$; and
\\
iv) $\func = \left( \funcEntity{\node{1}}, \dots, \funcEntity{\node{k}} \right) $ is a tuple of next-state multi-valued functions, such that if
$\nbhEntity{\node{i}} = \{\node{i_1} ,\ldots, \node{i_n} \}$ then the function
$\funcEntity{\node{i}} : \ssEntity{\node{i_1}} \times \dots \times \ssEntity{\node{i_n}} \rightarrow \ssEntity{\node{i}}$
defines the next state of $\node{i}$.
\pfbox
\\

\begin{figure}[b]
\centering
\begin{tabular}{c@{\qquad}c}
  \begin{tabular}[b]{c}
      \includegraphics[width=0.35\textwidth,keepaspectratio]{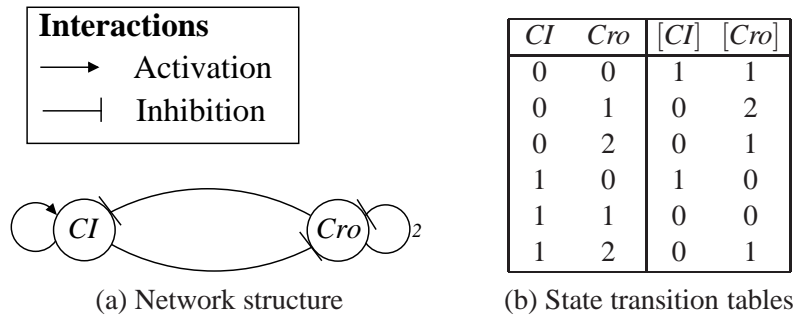}
      \\
   \end{tabular}

        &
      \begin{tabular}[b]{|cc|cc|}
        \hline $\CI$  & $\Cro$ & $\!\!\nextSt{\CI}\!\!$ & $\!\!\nextSt{\Cro}\!\!$\\
        \hline
        0 & 0 & 1 & 1\\
        0 & 1 & 0 & 2\\
        0 & 2 & 0 & 1\\
        1 & 0 & 1 & 0\\
        1 & 1 & 0 & 0\\
        1 & 2 & 0 & 1\\
        \hline
     \end{tabular}

   \\
  (a) Network structure & (b) State transition tables
\end{tabular}
\caption{
The MVN model $\PLI$ of the core regulatory mechanism
for the lysis-lysogeny switch in bacteriophage $\lambda$ (taken from \cite{Thieffry1995}).}
\label{fig:exPhage1}
\end{figure}

Consider the following simple example $\PLI$ of an MVN defined in Figure \ref{fig:exPhage1}
which models the core regulatory mechanism for the
{\it lysis--lysogeny switch} \cite{Thomas1990, opp05} in the bacteriophage $\lambda$
(this model is taken from \cite{Thieffry1995}).
%
It consists of two entities $\CI$ and $\Cro$, defined
such that $\ssEntity{\CI} = \{0,1\}$ and $\ssEntity{\Cro} = \{0,1,2\}$.
The next-state functions for each entity are defined using the state transition tables
presented in Figure \ref{fig:exPhage1}.(b) (where $\nextSt{\node{i}}$ is used to denote
the next state of entity $\node{i}$).
We can summarise the interactions as follows:
entity $\Cro$ inhibits the expression of $\CI$ and
at higher levels of expression, also inhibits itself;
entity $\CI$ inhibits the expression of $\Cro$ while
promoting its own expression.


In the sequel, let $\mvn = (\nodeAll, \ss, \nbh, \func)$ be an arbitrary MVN.
In a slight abuse of notation we let $\node{i} \in \mvn$ represent that $\node{i} \in \nodeAll$ is an entity in $\mvn$.

A {\it global state} of an MVN $\mvn$ with $k$ entities is
represented by a tuple of states $(s_{1}, \ldots, s_{k})$, where
$s_{i} \in \ssEntity{\node{i}}$ represents the state of entity
$\node{i} \in \mvn$. As a notational convenience we often use $s_{1}
\ldots s_{k}$ to represent a global state $(s_{1}, \ldots, s_{k})$.
When the current state of an MVN is clear from the context we let
$\node{i}$ denote both the name of an entity and its corresponding
current state. The {\it global state space} of an MVN $\mvn$,
denoted $\modelSS{\mvn}$, is the set of all possible global states
$\modelSS{\mvn}  = \ssEntity{\node{1}} \times \cdots \times
\ssEntity{\node{k}}$.

The state of an MVN can be updated either {\it synchronously} (see
\cite{kauffman1993,wuensch2002}), where the state of all entities is
updated simultaneously in a single update step, or {\it
asynchronously}\footnote{Note that different variations of the
asynchronous semantics  have been considered in the literature (see
for example \cite{Saadatpour2010}) but that we focus on the one most
commonly used for MVNs.} (see \cite{Thomas1973,Harvey1997}), where
entities update their state independently. We define these update
strategies more formally as follows:
\\
\\
\addDefn{Definition}{def:Update}
\\
1) {\it Synchronous Update}: Given two states $S_{1}, S_{2} \in
\modelSS{\mvn}$, we let $S_{1} \upStepSyn S_{2}$ represent a {\it
synchronous update step} such that $S_{2}$ is the state that results
from simultaneously updating the state of each entity $\node{i}$
using its next-state function $f_{\node{i}}$ and the appropriate
states from $S_{1}$ as indicated by the neighbourhood
$\nbhEntity{\node{i}}$.
\\
\\
2) {\it Asynchronous Update}: For any $\node{i} \in \mvn$ and any
state $S \in \modelSS{\mvn}$ we let $\upD{S}{\node{i}}$ denote the
global state that results by updating the state of $\node{i}$ in $S$
using $\funcEntity{\node{i}}$. Define the global state function
$\nxtState{\mvn} : \modelSS{\mvn} \rightarrow
\mathcal{P}(\modelSS{\mvn})$ on any state $S \in \modelSS{\mvn}$ by
$$\nxtState{\mvn}(S) = \{ \upD{S}{\node{i}} \ | \ \node{i} \in \mvn \ and \ \upD{S}{\node{i}} \not = S \}$$
Given a state $S_{1} \in \modelSS{\mvn}$ and $S_{2} \in
\nxtState{\mvn}(S_{1})$, we let $S_{1} \upStepAsy S_{2}$ represent
an {\it asynchronous update step}. 
\pfbox
\\

Note that given the above definition, only asynchronous update steps that result in a change in the current
state are considered (see \cite{Harvey1997}).

Continuing with our example, consider the global state $12$ for $\PLI$ (see Figure \ref{fig:exPhage1})
in which $\CI$ has state $1$ and $\Cro$ has state $2$.
Then
$12 \upStepSyn 01$
is a single synchronous update step on this state resulting in the new state $11$.
Considering an asynchronous update, we have
$\nxtState{\mvn}(12) = \{ 02, \ 11 \}$
and
$12 \upStepAsy 02$
and
$12 \upStepAsy 11$
are valid asynchronous update steps.

The sequence of update steps from an initial global state through
$\modelSS{\mvn}$ is called a {\it trace}. In the case of the
synchronous update semantics such traces are deterministic and
infinite. Given that the global state space is finite, this implies
that a synchronous trace must eventually enter a cycle, known
formally as an {\it attractor cycle} \cite{kauffman1993,Thomas1995}.
\\
\\
\addDefn{Definition}{def:synTrace} A {\it synchronous trace}
$\sigma$ is a list of global states $\sigma = \left\langle S_{0},
S_{1}, S_{2}, \dots \right\rangle$, where $S_{i} \upStepSyn
S_{i+1}$, for $i \geq 0$. \pfbox
\\

The set of all synchronous traces, denoted $\trSetS{\mvn}$,
therefore completely characterizes the behaviour of an MVN model
under the synchronous semantics and is referred to as the {\it
synchronous trace semantics} of $\mvn$. Note that we have one
synchronous trace for each possible initial state and so the set of
synchronous traces is always finite (see
\cite{kauffman1993,wuensch2002}).

In the asynchronous case, traces are non-deterministic and can be finite or infinite.
A single initial state can have an infinite number of possible asynchronous traces
starting from it and thus in the asynchronous case there can be infinite number of traces.
\\
\\
\addDefn{Definition}{def:asyTrace} An {\it asynchronous trace}
$\sigma$ is either:
\\
i) a finite sequence of global states
$\sigma = \left\langle S_{0}, S_{1}, \dots, S_{n} \right\rangle$,
where $S_{i} \upStepAsy  S_{i+1}$, for $i = 0,\ldots,n-1$,
and $\nxtState{\mvn}(S_{n}) = \{ \}$.
\\
ii) an infinite sequence of global states
$\sigma = \left\langle S_{0}, S_{1}, S_{2}, \dots \right\rangle$,
where $S_{i} \upStepAsy  S_{i+1}$, for $i \geq 0$.
\pfbox
\\

The set of all asynchronous traces, denoted $\trSetA{\mvn}$,
therefore completely characterizes the behaviour of an MVN model
under the asynchronous semantics and is referred to as the {\it
asynchronous trace semantics} of $\mvn$. Any state $S \in
\modelSS{\mvn}$  which cannot be asynchronously updated, i.e.
$\nxtState{\mvn}(S) = \{ \}$, is referred to as a {\it point
attractor} \cite{Thomas1990}. 

In our running example, $\PLI$ has a state space of size $|\modelSS{\PLI} | = 6$ and
has the following (finite in this case) set of asynchronous traces:
\begin{center}
\begin{tabular}[b]{l l}
     $\left\langle 00, 01, 02, 01, 02, \ldots  \right\rangle$ &
         $\left\langle 10  \right\rangle$ \\
     $\left\langle 00, 10  \right\rangle$ &
         $\left\langle 11, 01, 02, 01, 02, \ldots  \right\rangle$ \\
     $\left\langle 01, 02, 01, 02, \ldots  \right\rangle$ &
          $\left\langle 11, 10  \right\rangle$ \\
     $\left\langle 02, 01, 02, 01, \ldots  \right\rangle$ &
          $\left\langle 12, 02, 01, 02, 01, \ldots  \right\rangle$ \\
\end{tabular}
\end{center}
From the above traces it is clear that state $10$ is a point attractor for $\PLI$.

The behaviour of an MVN under the synchronous or asynchronous trace semantics can be represented by a
state graph (for example, see \cite{Tournier2009})
in which the nodes are the global states and the edges are precisely the update steps allowed.
We let
$\stGraphSyn{\mvn} = (\modelSS{\mvn}, \upStepSyn)$
and
$\stGraphAsy{\mvn} = (\modelSS{\mvn}, \upStepAsy)$
denote the corresponding state graphs under the synchronous and asynchronous trace semantics.

\begin{figure}[tb]
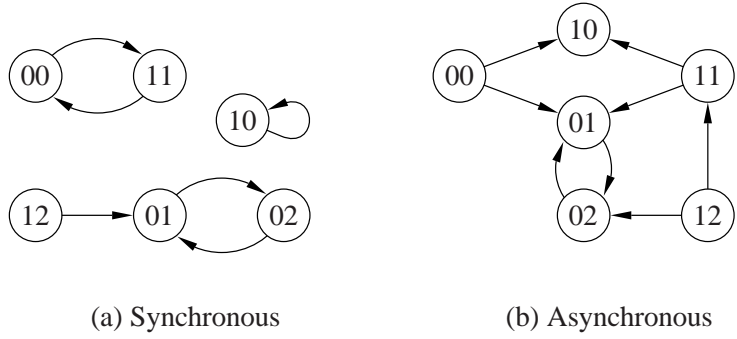

\centering
\begin{tabular}[b]{c@{\qquad}c@{\qquad}c}
\begin{graph}(4,4)(0,-1)
\graphnodecolour{1}
\graphnodesize{1.3}
\grapharrowlength{0.5}
\graphlinewidth{0.03}
\grapharrowwidth{0.4}

\NonStandardSize

\stateNode{s00}{0}{3.375}{00}
\stateNode{s01}{3}{0}{01}
\stateNode{s02}{6}{0}{02}
\stateNode{s10}{5}{2.3}{10}
\stateNode{s11}{3}{3.375}{11}
\stateNode{s12}{0}{0}{12}

\dirbow{s00}{s11}{0.3}
\dirbow{s11}{s00}{0.3}
\dirloopedge{s10}{45}(1,0)
\diredge{s12}{s01}
\dirbow{s01}{s02}{0.3}
\dirbow{s02}{s01}{0.3}

\end{graph}

        & \ \ &

\begin{graph}(4,4)(0,-1)
\graphnodecolour{1}
\graphnodesize{1.3}
\grapharrowlength{0.5}
\graphlinewidth{0.03}
\grapharrowwidth{0.4}

\NonStandardSize

\stateNode{s00}{0}{3.375}{00}
\stateNode{s01}{3}{2.25}{01}
\stateNode{s02}{3}{0}{02}
\stateNode{s10}{3}{4.5}{10}
\stateNode{s11}{6}{3.375}{11}
\stateNode{s12}{6}{0}{12}

\diredge{s00}{s01}
\diredge{s00}{s10}
\diredge{s11}{s01}
\diredge{s11}{s10}
\diredge{s12}{s02}
\diredge{s12}{s11}
\dirbow{s01}{s02}{0.3}
\dirbow{s02}{s01}{0.3}

\end{graph}

\\
  (a) Synchronous & & (b) Asynchronous \\
\end{tabular}
\caption
    {The (a) synchronous and (b) asynchronous state graphs for $\PLI$.
    }
\label{figStateGraphsPLI}
\end{figure}

The synchronous and asynchronous state graphs for  $\PLI$ are
presented in Figure \ref{figStateGraphsPLI}.

When analysing the behaviour of an MVN it is important to consider its attractors which can
represent important biological phenomena, such as
different cellular types like proliferation, apoptosis and differentiation \cite{huang2000}.
In the synchronous case all traces are infinite and so must lead to
a cyclic sequence of states which are taken as an {\it attractor}
\cite{kauffman1993,Thomas1995,wuensch2002}. As an example, consider
$\PLI$ (see Figure \ref{figStateGraphsPLI}.(a)) which has the point
attractor $10 \updateStep 10$; and attractors $00 \upStepSyn 11
\upStepSyn 00$  and $01 \upStepSyn 02 \upStepSyn 01$ of period~2. In
the asynchronous case we have {\it point attractors} which are
states that cannot be updated and also the strongly connected
components in an MVN's asynchronous state graph are considered to be
{\it attractors}~\cite{Tournier2009}. Again, considering $\PLI$ (see
Figure \ref{figStateGraphsPLI}.(b)) we can see that in the
asynchronous case it has two point attractors, $01$ and $10$, and
one attractor $01 \upStepAsy 02 \upStepAsy 01$.

\section{Asynchronous Abstractions}
\label{sec:asynAbs}
In this section we consider developing a notion of abstraction for asynchronous MVNs.
The idea is to formulate what it means for an MVN to be correctly abstracted by a simpler MVN
with the same network structure but smaller state space.
We take as our starting point the abstraction techniques developed for synchronous MVNs \cite{banksSteggles2010}
and investigate extending these to the asynchronous case.
We show that our approach allows sound analysis inferences about positive reachability properties and
that all attractors of an asynchronous abstraction
correspond to attractors in the original MVN.

We begin by recalling the notion of a state mapping and abstraction mapping \cite{banksSteggles2010}
used to reduce an entity's state space.
\\
\\
\addDefn{Definition}{def:sm} Let $\mvn$ be an MVN and let $\node{i}
\in \mvn$ be an entity such that $\ss(\node{i}) = \{0, \ldots, m \}$
for some $m > 1$. Then a {\it state mapping} $\mapping{\node{i}}$
for entity $\node{i}$ is a surjective mapping $\mapping{\node{i}} :
\{0, \ldots, m \} \rightarrow\ \{0, \ldots,n \}$, where $0 < n < m$.
\pfbox
\\

The state mapping must be surjective to ensure that all states in the new reduced state space are used.
From a biological viewpoint it may also be reasonable to further restrict the state mappings considered,
for example, only considering those mappings which are order-preserving.
Note we only consider state mappings with a codomain larger than one, since a singular state
entity does not appear to be of biological interest.

As an example, consider entity $\Cro \in \PLI$ (see Figure \ref{fig:exPhage1}) which has the state space
$\ssEntity{\Cro} = \{0, 1, 2 \}$.
It is only meaningful to simplify $\Cro$ to a Boolean entity and so
one possible state mapping to achieve this would be:
$$
\mapping{\Cro} = \{0 \mapsto 0, 1 \mapsto 1, 2 \mapsto 1\},
$$
which maps state $0$ to $0$ and merges states $1$ and $2$ into a single state $1$.

In order to be able to simplify several entities at the same time during the abstraction process
we introduce the notion of a family of state mappings.
\\
\\
\addDefn{Definition}{def:absMap} Let $\mvn = (\nodeAll, \ss, \nbh,
\func)$ be an MVN with entities $G = \{ \node{1} , \dots, \node{k}
\}$. Then an {\it abstraction mapping} $\absMap = \langle
\mapping{\node{1}}, \ldots, \mapping{\node{k}} \rangle$ for $\mvn$
is a family of mappings such that for each $1 \leq i \leq k$ we have
$\mapping{\node{i}}$ is either a state mapping for entity $\node{i}$
or is the identity mapping $I_{\node{i}} : \ss(\node{i}) \rightarrow
\ss(\node{i})$ where $I_{\node{i}}(s) = s$, for all $s \in
\ss(\node{i})$. Furthermore, for $\absMap$ to be useful we normally
insist that at least one of the mappings $\mapping{\node{i}}$ is a
state mapping. \pfbox
\\

Note in the sequel given a state mapping $\mapping{\node{i}}$ we let it denote both itself and the
corresponding abstraction mapping containing
only the single state mapping $\mapping{\node{i}}$.

An abstraction mapping $\absMap$ can be used to abstract an asynchronous trace (see Definition \ref{def:asyTrace}) using a
similar approach to that detailed for synchronous traces \cite{banksSteggles2010}.
We begin by defining how an abstraction mapping can be lifted to a global state.
\\
\\
\addDefn{Definition}{dfn:absState}
Let $\absMap = \langle \mapping{\node{1}} \ldots \mapping{\node{k}} \rangle$
be an abstraction mapping for $\mvn$.
Then $\absMap$ can be used to abstract a global state
$s_{1} \ldots s_{k} \in \modelSS{\mvn}$
by applying it pointwise, i.e.
$\absMap(s_{1}\dots s_{k}) = \mapping{\node{1}}(s_{1}) \ldots \mapping{\node{k}} (s_{k})$.
\pfbox
\\

We can apply an abstraction mapping $\absMap$ to an asynchronous trace $\sigma \in \trSetA{\mvn}$
by applying $\absMap$ to each global state in the trace in the obvious way and
and then merging consecutive identical states.
Note that removing consecutive identical states is needed since by the definition of an
asynchronous trace (see Definition \ref{def:asyTrace}) each asynchronous update rule must result in a new global state,
i.e. the state of an entity has to change in order for a state transition to occur.
\\
\\
\addDefn{Definition}{dfn:absAsyTrace}
Let $\absMap = \langle \mapping{\node{1}} \ldots \mapping{\node{k}} \rangle$
be an abstraction mapping for $\mvn$ and let
$\sigma  \in \trSetA{\mvn}$
be either a finite
$\sigma = \left\langle S_{0}, S_{1}, \dots, S_{n} \right\rangle$
or infinite
$\sigma = \left\langle S_{0}, S_{1}, S_{2}, \dots \right\rangle$
asynchronous trace.
Then $\absMap(\sigma)$ is the abstracted trace that results by
\\
\\
i) First apply the abstraction mapping to each state in $\sigma$, i.e.
in the finite case
$\left\langle \absMap(S_{0}), \absMap(S_{1}), \dots, \absMap(S_{n}) \right\rangle$
or in the infinite case
$\left\langle \absMap(S_{0}), \absMap(S_{1}), \absMap(S_{2}), \dots \right\rangle$.
\\
ii) Next merge consecutive identical global states in the trace into a single global state
to ensure that no two consecutive states are identical in the resulting abstracted trace, i.e.
suppose the result is an infinite trace
$\left\langle \absMap(S_{0}), \absMap(S_{1}), \absMap(S_{2}), \dots \right\rangle$
then we know that for $i \in \nat$ we have $\absMap(S_{i}) \not = \absMap(S_{i+1})$.
\pfbox
\\

We let
$\absMap(\trSetA{\mvn}) = \{\absMap(\sigma) \ | \ \sigma \in \trSetA{\mvn} \}$
denote the set of abstracted traces.

As an example, consider applying the abstraction mapping
$
\mapping{\Cro} = \{0 \mapsto 0, 1 \mapsto 1, 2 \mapsto 1\}
$
to the $\PLI$ asynchronous trace
$\left\langle 00, 01, 02, 01, 02, \ldots  \right\rangle$.
Part i) of Definition \ref{dfn:absAsyTrace} above results in the trace
$\left\langle 00, 01, 01, 01, 01, \ldots  \right\rangle$;
we now merge identical consecutive states to derive the abstracted trace
$\left\langle 00, 01  \right\rangle$.
It is interesting to note that abstracting an infinite trace can result in a finite
abstracted trace, as above.
The intuition here is that a cyclic set of states have been abstracted to a single point.
The complete set of abstracted asynchronous traces of $\PLI$ using $\mapping{\Cro}$ are given below:
\begin{center}
\begin{tabular}[b]{l c l}
     $\mapping{\Cro}(\left\langle 00, 01, 02, 01, 02, \ldots  \right\rangle)$ & $=$ & $\left\langle 00, 01,   \right\rangle$ \\
     $\mapping{\Cro}(\left\langle 00, 10  \right\rangle)$ & $=$ & $\left\langle 00, 10  \right\rangle$ \\
     $\mapping{\Cro}(\left\langle 01, 02, 01, 02, \ldots  \right\rangle)$ & $=$ & $\left\langle 01   \right\rangle$ \\
     $\mapping{\Cro}(\left\langle 02, 01, 02, 01, \ldots  \right\rangle)$ & $=$ & $\left\langle 01  \right\rangle$ \\
     $\mapping{\Cro}(\left\langle 10  \right\rangle)$ & $=$ & $\left\langle 10  \right\rangle$ \\
     $\mapping{\Cro}(\left\langle 11, 01, 02, 01, 02, \ldots  \right\rangle)$ & $=$ & $\left\langle 11, 01  \right\rangle$ \\
     $\mapping{\Cro}(\left\langle 11, 10  \right\rangle)$ & $=$ & $\left\langle 11, 10  \right\rangle$ \\
     $\mapping{\Cro}(\left\langle 12, 02, 01, 02, 01, \ldots  \right\rangle)$ & $=$ & $\left\langle 11, 01 \right\rangle$ \\
\end{tabular}
\end{center}


The definition of an asynchronous abstraction is based on its trace
semantics and follows along similar lines to that for the
synchronous case \cite{banksSteggles2010}. We say an asynchronous
abstraction is correct if its set of traces is within the abstracted
traces of the original MVN. This definition of abstraction
represents an {\it under--approximation} \cite{} since not all of
the behaviour of the original MVN is guaranteed to have been
captured within the abstraction (we discuss the implications of this
below).
\\
\\
\addDefn{Definition}{def:asynAbs} Let $\mvni{1} = (\nodeAll_{1},
\ss_{1}, \nbh_{1}, \func_{1})$ and $\mvni{2} = (\nodeAll_{2},
\ss_{2}, \nbh_{2}, \func_{2})$ be two MVNs with the same structure,
i.e. $\nodeAll_{1} = \nodeAll_{2}$ and $\nbh_{1}(\node{i}) =
\nbh_{2}(\node{i})$, for all $\node{i} \in \mvni{1}$. Let $\absMap$
be an abstraction mapping from $\mvni{2}$ to $\mvni{1}$. Then we say
that $\mvni{1}$ {\it asynchronously abstracts} $\mvni{2}$ {\it
under} $\absMap$, denoted $\mvni{1} \absA{\absMap} \mvni{2}$, if,
and only if, $\trSetA{\mvni{1}} \subseteq
\absMap(\trSetA{\mvni{2}})$. \pfbox
\\

As an abstraction example, consider the MVN  $\PLIab$ defined in Figure \ref{fig:abstractPhage1}
which has the same structure as $\PLI$ (see Figure \ref{fig:exPhage1}) but is a Boolean model.
\begin{figure}[h]
\centering
\begin{tabular}[b]{c@{\qquad}c}

      \begin{tabular}[b]{|cc|cc|}
        \hline $\CI$  & $\Cro$ & $\!\!\nextSt{\CI}\!\!$ & $\!\!\nextSt{\Cro}\!\!$\\
        \hline
        0 & 0 & 1 & 1\\
        0 & 1 & 0 & 1\\
        1 & 0 & 1 & 0\\
        1 & 1 & 0 & 0\\
        \hline
     \end{tabular}

   &
   \begin{tabular}[b]{l l}
      $\left\langle 00, 01  \right\rangle$ &
        $\left\langle 10  \right\rangle$ \\
      $\left\langle 00, 10  \right\rangle$ &
        $\left\langle 11, 01  \right\rangle$ \\
      $\left\langle 01  \right\rangle$ &
        $\left\langle 11, 10  \right\rangle$ \\
      \\
   \end{tabular}
\end{tabular}
\caption{
State transition tables defining $\PLIab$ and associated asynchronous trace semantics $\trSetA{\PLIab}$.}
\label{fig:abstractPhage1}
\end{figure}
Then given the abstraction mapping
$\mapping{\Cro} = \{0 \mapsto 0, 1 \mapsto 1, 2 \mapsto 1\}$
we can see that
$\trSetA{\PLIab} \subseteq \mapping{\Cro}(\trSetA{\PLI})$
holds and so $\PLIab$ is an abstraction of $\PLI$,
i.e. $\PLIab \absA{\mapping{\Cro}} \PLI$ holds.
Note that $\PLIab$ has two point attractors:
$01$ and $10$ which correspond to the two attractors associated with $\PLI$
(see Figure \ref{figStateGraphsPLI}.(b))
and thus, $\PLIab$ can bee seen to be a good approximation of the behaviour of $\PLI$.

Recall that one of the original motivations for developing an
abstraction theory was to aid
the analysis of complex MVNs.
It is therefore important to consider what properties of an asynchronous MVN can be
inferred from an abstraction MVN.
We consider reachability and the existence of
attractors since these are the main properties that are considered when analysing
an MVN.
\\
\\
\addDefn{Theorem}{thm:reachability}
Let $\mvni{1} \absA{\absMap} \mvni{2}$
and let $S_{1}, S_{2} \in \modelSS{\mvni{1}}$.
If $S_{2}$ is reachable from $S_{1}$ in $\mvni{1}$ then there must exist states
$S_{1}', S_{2}' \in \modelSS{\mvni{2}}$ such that
$\absMap(S_{1}') = S_{1}$,
$\absMap(S_{2}') = S_{2}$, and
$S_{2}'$ is reachable from $S_{1}'$ in $\mvni{2}$.
\\
\\
\textbf{Proof.} Since $S_{2}$ is reachable from $S_{1}$ there must
exist a trace $\sigma \in \trSetA{\mvni{1}}$ which begins with
state~$S_{1}$ and which contains state $S_{2}$. From Definition
\ref{def:asynAbs}, we know that $\trSetA{\mvni{1}} \subseteq
\absMap(\trSetA{\mvni{2}})$ must hold. Therefore there must exist a
trace $\sigma' \in \trSetA{\mvni{2}}$ such that $\absMap(\sigma') =
\sigma$. From this it is straightforward to see that there must
exist the required states $S_{1}'$ and $S_{2}'$ in $\sigma'$ such
that $\absMap(S_{1}') = S_{1}$, $\absMap(S_{2}') = S_{2}$, and
$S_{2}'$ is reachable from $S_{1}'$. \pfbox
\\

The above theorem indicates that inferring reachability properties from an abstraction is
sound but not complete \cite{DSilva2008}.
The implications of this can be summarised as follows:
(i) If one state is reachable from another in an abstraction then a corresponding
reachability property must hold in the original model;
(ii) However, if one state is not reachable from another in an abstraction then a corresponding
reachability property in the original MVN may or may not hold and more analysis will be required.
This relates to the fact that our notion of abstraction represents
an {\it under--approximation} \cite{Clarke1994,Pelanek2006} of the
original model. The alternative approach would be to use an {\it
over--approximation} abstraction model
\cite{Clarke1994,Pelanek2006,Clarke03} in which false positives can
arise and need to be dealt with. It turns out that an
over--approximation approach is not well suited to MVNs given that
our goal is to find an abstraction model that is a well--defined
MVN. To illustrate the potential problems, consider what happens if
a point attractor is identified to a non--attractor state by an
abstraction mapping. In this case no over--approximation abstraction
can exist since such an MVN would need to contain a state that was
both a point attractor and also had a successor state. Thus the
approach taken here of using an under--approximation appears to be
the appropriate approach to use.

Note that a consequence of the above is that all attractors in an abstraction must have
corresponding attractors in the original MVN.
\\
\\
\addDefn{Corollary}{corr:attractors}
If $\mvni{1} \absA{\absMap} \mvni{2}$
then all attractors of $\mvni{1}$ must represent attractors
in $\mvni{2}$.
\\
\\
\textbf{Proof.} Follows directly from the definition of an attractor and
Theorem \ref{thm:reachability}.
\pfbox
\\

\section{A Decision Procedure for Asynchronous Abstractions}
\label{sec:decProc}
Given we have now formulated a definition of an asynchronous abstraction we are now interested in defining a procedure for
checking whether a proposed abstraction
$\mvni{1}$ is an asynchronous abstraction of an MVN $\mvni{2}$.
In the synchronous case the approach taken was to simply check that each trace $\sigma \in \trSetS{\mvni{1}}$ was
contained within the set of abstracted traces $\absMap(\trSetS{\mvni{2}})$.
However, in the asynchronous case both sets of traces $\trSetA{\mvni{1}}$ and $\absMap(\trSetA{\mvni{2}})$
may be infinite and so such a simple set inclusion check is not feasible.
Instead we propose a decision procedure based on using the state graphs that summarise
the behaviour of an asynchronous MVN.
The idea is to consider all sets of states and associated edges that can be used to model an abstract state.
We then iterate through these removing those state sets which can not be represented given the current allowable state sets.
If at any point we no longer have any state sets remaining for a particular abstract state then we have shown the abstraction is
not valid and we terminate the decision procedure.
If, on the other hand, we reach a point at which no more state sets can be removed then we know the abstraction must be valid and we can again terminate the procedure.

In the sequel let $\mvni{1}$ and $\mvni{2}$ be MVNs with the same structure
and let $\absMap$ be an abstraction mapping from $\mvni{2}$ to $\mvni{1}$.

In order to define a decision procedure
{\tt checkAsynAbs($MV_{1}$,$MV_{2}$,$\absMap$)}
for checking if $\mvni{1}$ is
an asynchronous abstraction under $\absMap$ of $\mvni{2}$
we begin by formulating some preliminary concepts.
\\
\\
i) {\it Representing abstract states}: Let $S \in
\ssEntity{\mvni{1}}$ then we define
$$\setStAbs{S} = \{ S' \ | \ S' \in \ssEntity{\mvni{2}}, \ \absMap(S') = S \}$$
to be the set of all states in $\mvni{2}$ that can represent the abstract state $S$.
\\
\\
ii) {\it Set of identical consecutive states}: For any state $S' \in
\ssEntity{\mvni{2}}$ we define the set $\equivClass{S'}$ of all {\it
consecutive} reachable states from $S'$ that have the same abstract
state $\absMap(S')$. Define 
$\equivClass{S'} = \bigcup_{i \in \nat} \equivClass{S'}_{i}$,
where
$\equivClass{S'}_{i}$
is defined recursively:
$\equivClass{S'}_{0} = \{ S' \}$
and
$$\equivClass{S'}_{i+1} = \{ S'_{2} \ | \ S'_{1} \in \equivClass{S'}_{i},
\ S'_{2} \in \nxtState{\mvni{2}}(S'_{1}), \ \absMap(S') = \absMap(S'_{2}) \}.$$

We now define the notion of a {\it step term}, an expression which
is used to represent one possible way to model an abstract state
using a set of original states. Such step terms will form the basis
of our decision procedure.
\\
\\
\addDefn{Definition}{dfn:stepTerm} Let $S \in \ssEntity{\mvni{1}}$
and suppose $\nxtState{\mvni{1}}(S) = \{ S_{1}, \ldots, S_{m} \}$.
Then for each non-empty set of states $\Gamma \subseteq
\setStAbs{S}$ we define the {\it step term} $\stShort{\Gamma}{S}$ by
$$\stShort{\Gamma}{S} = \stepTerm{S}{\Gamma}{\setS{S_{1}}, \ldots, \setS{S_{m}}}, $$
where
$\setS{S_{i}} = \{ S' _{2} \ | \ S'_{1} \in \Gamma, \
S'_{2} \in \nxtState{\mvni{2}}(\{ S'_{1} \} \ \cup \equivClass{S'_{1}}),  \
\absMap(S'_{2}) = S_{i} \},$
and $\nxtState{\mvni{2}}$ has been lifted from taking a single state as input to taking a set of states
in the obvious way.
Note that the use of $\equivClass{S'_{i}}$ is needed in the above definition to take
account of the merging of consecutive identical states that occurs in
abstracted traces (see part ii) in Definition \ref{dfn:absAsyTrace}).

We say a step term $\stepTerm{S}{\Gamma}{\setS{S_{1}}, \ldots,
\setS{S_{m}}}$ is {\it valid} iff:
\\
i) the states $\Gamma$ used in a step term have the appropriate connections, i.e.
$\setS{S_{i}} \not = \{ \}$, for $i=1, \ldots, m$; and
\\
ii) if $S$ is a point attractor in $\stGraphAsy{\mvni{1}}$ then it must be modelled by point attractors in
$\stGraphAsy{\mvni{2}}$ (discounting steps to identical abstracted states), i.e. if $\nxtState{\mvni{1}}(S) = \{ \}$
then for each $S' \in \Gamma$
we have
$\nxtState{\mvni{2}}(\equivClass{S'}) - \equivClass{S'} = \{ \}$.
\pfbox
\\

We let $\allStepTerm{S}$
denote the set of all valid step terms
$$\allStepTerm{S} = \{ \stShort{\Gamma}{S} \ | \ \Gamma \subseteq \setStAbs{S}, \ \stShort{\Gamma}{S} \ {\rm {\it is \ valid}} \}. $$

Observe that each valid step term
$\stShort{\Gamma}{S} \in \allStepTerm{S}$
must correctly model in $\mvni{2}$ the connections between $S \in \ssEntity{\mvni{1}}$
and its corresponding next states
$\nxtState{\mvni{1}}(S)$ in $\mvni{1}$.

The proposed decision procedure is presented in Figure \ref{fig:decProc}.
It works by creating a family
$C = \langle C(S) \subseteq \allStepTerm{S} \ | \ S \in \ssEntity{\mvni{1}} \rangle$
of sets of all valid step terms.
It then repeatedly looks at each set of step terms $C(S)$,
for each abstract state
$S \in \ssEntity{\mvni{1}}$,
removing those that have next states that
are not currently in the remaining stored step terms of $C$.

\begin{figure}[h]
{\tt
\begin{center}
\begin{tabbing}
Algorithm checkAsynAbs($MV_{1}$,$MV_{2}$,$\absMap$): \\
\\
/** Initialise valid state terms **/ \\
for  \= each $S \in \ssEntity{MV_{1}}$  do
   $C(S) = \allStepTerm{S}$ \\
/** Iteratively check sets of step terms **/ \\
repeat \\
\> done:=true \\
\> for \= each $S \in \ssEntity{MV_{1}}$  do \\
\> \>for \= each
     $\stepTerm{S}{\Gamma}{\setS{S_{1}}, \ldots, \setS{S_{m}}} \in C(S)$  do \\
\> \> \>for \= $i$:= $1$ to $m$ do  \\
\> \> \> \>if \= $\stShort{\setS{S_{i}}}{S_{i}} \not \in C(S_{i})$ then \\
\> \> \> \> \>$C(S) = C(S) - \{ \stepTerm{S}{\Gamma}{\setS{S_{1}}, \ldots, \setS{S_{m}}} \}$ \\
\> \> \> \> \>done:=false \\
\> \> \>if $C(S) = \{ \}$ then return false\\
until (done = true) \\
return true
\end{tabbing}
\end{center}
}
\caption{
Decision procedure for checking asynchronous abstractions $\mvni{1} \absA{\absMap} \mvni{2}$.}
\label{fig:decProc}
\end{figure}

It is straightforward to show that the decision procedure must always terminate.
\\
\\
\addDefn{Theorem}{thm:decProTerminate}
The decision procedure
{\tt checkAsynAbs($MV_{1}$,$MV_{2}$,$\absMap$)}
always terminates.
\\ \\
\textbf{Proof.}
This follows from that fact we can only ever begin with a
finite family of finite sets of step terms, that no step terms can
ever be added, and that we must remove
at least on step term in order to continue to the next iteration.
Therefore the algorithm either terminates when no step terms are removed
or continues to iterate until we reach a point where one set
$C(S)$ of step terms is empty, again resulting in termination
of the algorithm.
\pfbox
\\

The complexity of the decision procedure in the worst case, when $\mvni{1}$
is not an asynchronous abstraction of $\mvni{2}$, can be derived as follows.
Assume $\mvni{1}$ is a Boolean model which has $n$ entities
and $k$ is an upper bound on the number of states in $\mvni{2}$
that can be abstracted to a single state in $\mvni{1}$,
i.e. $k \geq |\setStAbs{S}|$, for all $S \in \ssEntity{\mvni{1}}$.
Note that $k$ can be calculated from the abstraction mapping used and is not
dependent on $n$.
The three nested for loops in the decision procedure have an upper bound of
$O(2^n \times 2^k \times n)$
where:
$2^n$ is the number of states in $\mvni{1}$;
$2^k$ is an upper bound on the number of different sets of states that can be mapped
to a given abstract state; and
$n$ represents the maximum number of states that can be connected to a given state.
The outer repeat until loop will iterate round removing a single step term until
one of the step term sets is empty.
This gives a final upperbound of
$O(2^{2(n+k)} \times n^{2})$.
In practice the decision procedure should perform much better than this.
Note that for a given abstraction mapping, $k$ can be seen as a fixed constant which does not increase as entities are added (providing the state of those entities is not abstracted).

Let $\stepTerm{S}{\Gamma}{\setS{S_{1}}, \ldots, \setS{S_{m}}}$ be a valid step term,
let $\alpha_{1} \in \Gamma$ and $\alpha_{2} \in \setS{S_{i}}$, for some $1 \leq i \leq m$.
Then note that due to the way consecutive identical states are treated it may not directly hold that
$\alpha_{1} \upStepAsy \alpha_{2}$ since
$\alpha_{2} \in \nxtState{\mvni{2}}(\{ \alpha_{1} \} \ \cup \equivClass{\alpha_{1}})$.
We let
$\overline{\alpha_{1}} = \alpha_{1} \upStepAsy \alpha_{1}^{1} \upStepAsy \cdots \upStepAsy \alpha_{1}^{r}$,
for $\alpha_{1}^{j} \in \equivClass{\alpha_{1}}$, for $1 \leq j \leq r$ represent the sequence of identical
abstracted states needed such that
$\overline{\alpha_{1}} \upStepAsy \alpha_{2}$
does hold in $\mvni{2}$.

The following lemma considers how step terms can be chained together
and is is needed to prove the main correctness result below.
\\
\\
\addDefn{Lemma}{lem:allPaths}
Let $C = \langle C(S) \subseteq \allStepTerm{S} \ | \ S \in \ssEntity{\mvni{1}} \rangle$
be a family of sets of valid step terms such that:
\\
i) For each $S \in \ssEntity{\mvni{1}}$ we have
$C(S) \not = \{ \}$;
\\
ii) The family $C$ is closed under step terms, i.e.
for each $S \in \ssEntity{\mvni{1}}$ and
$\stepTerm{S}{\Gamma}{\setS{S_{1}}, \ldots, \setS{S_{m}}} \in C(S)$
we have
$\stShort{\setS{S_{i}}}{S_{i}} \in C(S_{i})$,
for  $1 \leq i \leq m$.
\\
\\
Then every  path\footnote{We note that a path differs from a trace in that a trace represents a complete run
of an MVN whereas a path is simply a walk through an MVN's state graph.}
$\gamma = \gamma_{1} \upStepAsy \ldots \upStepAsy \gamma_{p}$
in the abstraction state graph
$\stGraphAsy{\mvni{1}}$
must have a corresponding path
$\alpha = \alpha_{1} \upStepAsy \ldots \upStepAsy \alpha_{r}$,
$r \geq p$,
in the original state graph
$\stGraphAsy{\mvni{2}}$
such that
$\absMap(\alpha) = \gamma$.
\\
\\
\textbf{Proof.}
\\
Let $\gamma = \gamma_{1} \upStepAsy \cdots \upStepAsy \gamma_{p}$
be a path in the state graph
$\stGraphAsy{\mvni{1}}$.
Then by assumptions i) and ii) it is straightforward to see
there must exist a (not necessarily unique) chain of step terms
$$\stepTerm{\gamma_{i}}{\Gamma_{i}}{\ldots, \setS{\gamma_{i+1}}, \ldots} \in C(\gamma_{i}), \ \
\stShort{\Gamma_{p}}{\gamma_{p}} \in C(\gamma_{p})$$
for $1 \leq i < p$,
such that for $j=2,\ldots,p$ we have
$\Gamma_{j} = \setS{\gamma_{j}}$.

We now prove that for any $\alpha_{p} \in \Gamma_{p}$
there must exist $\alpha_{i} \in \Gamma_{i}$,
for $1 \leq i < p$, such that
$\alpha = \overline{\alpha_{1}} \upStepAsy \cdots \upStepAsy \overline{\alpha_{p-1}} \upStepAsy \alpha_{p}$
is a path in $\stGraphAsy{\mvni{2}}$
with
$\absMap(\alpha) = \gamma$.
We prove this using induction on $p \in \nat$, $p \geq 2$ as follows.
\\
\\
1) {\it Induction Base}. Let $p = 2$ and suppose we have a path
$\gamma_{1} \upStepAsy \gamma_{2}$. Then we know there must exist
step terms $\stepTerm{\gamma_{1}}{\Gamma_{1}}{\ldots,
\setS{\gamma_{2}}, \ldots} \in C(\gamma_{1})$ and
$\stShort{\setS{\gamma_{2}}}{\gamma_{2}} \in C(\gamma_{2})$ (as
explained above). Clearly by the definition of step terms we know
that for any $\alpha_{2} \in \setS{\gamma_{2}}$ there must exist
$\alpha_{1} \in \Gamma_{1}$ such that $\overline{\alpha_{1}}
\upStepAsy \alpha_{2}$ and
$$\absMap(\overline{\alpha_{1}} \upStepAsy \alpha_{2}) =  \gamma_{1} \upStepAsy \gamma_{2}.$$
2) {\it Induction Step}. Let $p=q+1$, for some $q \in \nat$, $q \geq
2$. Suppose we have a path $\gamma = \gamma_{1} \upStepAsy \cdots
\upStepAsy \gamma_{q} \upStepAsy \gamma_{q+1}$. Then we know there
must exist step terms
$$\stepTerm{\gamma_{1}}{\Gamma_{1}}{\ldots, \setS{\gamma_{2}}, \ldots} \in C(\gamma_{1}), \ \
  \stepTerm{\gamma_{i}}{\setS{\gamma_{i}}}{\ldots, \setS{\gamma_{i+1}}, \ldots} \in C(\gamma_{i}), \ \
  \stShort{\setS{\gamma_{q+1}}}{\gamma_{q+1}} \in C(\gamma_{q+1}),
$$
for $2 \leq i \leq q$ (as explained above).
Then by the induction hypothesis we know for each $\alpha_{q} \in \setS{\gamma_{q}}$ there must exist
$\alpha_{i} \in \Gamma_{i}$, for $1 \leq i < q$,
such that
$\overline{\alpha_{1}} \upStepAsy \cdots \upStepAsy \overline{\alpha_{q-1}} \upStepAsy \alpha_{q}$
is a path in $\stGraphAsy{\mvni{2}}$
with
$\absMap(\overline{\alpha_{1}} \upStepAsy \cdots \upStepAsy \overline{\alpha_{q-1}} \upStepAsy \alpha_{q})
= \gamma_{1} \upStepAsy \cdots \upStepAsy \gamma_{q}$.
By the definition of step terms it follows that for any
$\alpha_{q+1} \in \setS{\gamma_{q+1}}$
there must exist
$\alpha_{q} \in \Gamma_{q}$
such that
$\overline{\alpha_{q}} \upStepAsy \alpha_{q+1}$.
Combining this with the induction hypothesis given above shows
the existence of the required path in $\stGraphAsy{\mvni{2}}$.
\pfbox \\

It now remains to show that the decision procedure {\tt checkAsynAbs($MV_{1}$,$MV_{2}$,$\absMap$)}
correctly checks for asynchronous abstractions.
\\
\\
\addDefn{Theorem}{thm:decProCorrect}
{\tt checkAsynAbs($MV_{1}$,$MV_{2}$,$\absMap$)}
returns $true$
if, and only if,
$\mvni{1} \absA{\absMap} \mvni{2}$.
\\
\\
\textbf{Proof.}
\\
{\it Part 1)} $\Rightarrow$ Suppose {\tt
checkAsynAbs($MV_{1}$,$MV_{2}$,$\absMap$)} returns $true$. By
inspecting the decision procedure we can see this means that a
family $\{ C(S)\subseteq \allStepTerm{S} \ | \ S \in
\ssEntity{\mvni{1}} \}$ of non--empty sets of valid step terms must
have been found which is closed under step terms. 
Consider any abstract trace $\sigma \in \trSetA{\mvni{1}}$; then by
Lemma \ref{lem:allPaths} and since any trace can be interpreted as a
path in $\stGraphAsy{\mvni{1}}$ we have that there must exist a path
$\alpha$ in $\stGraphAsy{\mvni{2}}$ such that $\absMap(\alpha) =
\sigma$. It is straightforward to see that $\alpha$ must be a
well--defined trace for $\mvni{2}$, i.e. $\alpha \in
\trSetA{\mvni{2}}$, by the definition of {\it valid} step term. 
This shows that
$\trSetA{\mvni{1}} \subseteq \absMap(\trSetA{\mvni{2}})$
and so by Definition \ref{def:asynAbs} we have
$\mvni{1} \absA{\absMap} \mvni{2}$.
\\
\\
{\it Part 2)} $\Leftarrow$ Suppose $\mvni{1} \absA{\absMap}
\mvni{2}$ then by Definition \ref{def:asynAbs} we know
$$\trSetA{\mvni{1}} \subseteq \absMap(\trSetA{\mvni{2}}) \eqno{(1)}$$
Then we show that there must exist a family of sets of valid step terms which
are closed under step term inclusion and thus that
{\tt checkAsynAbs($MV_{1}$,$MV_{2}$,$\absMap$)}
must terminate returning $true$.

Let $X \subseteq \trSetA{\mvni{2}}$ be the set of traces that abstractly correspond to $\trSetA{\mvni{1}}$:
$$X = \{ \sigma \ | \ \sigma' \in \trSetA{\mvni{2}}, \ \
    \exists \sigma \in \trSetA{\mvni{1}} . \absMap(\sigma') = \sigma \}
$$
For each $S \in \ssEntity{\mvni{1}}$, let $X\langle S \rangle$
denote the set of all states
that abstract to $S$ which occur at the start of a trace in $X$:
$$X\langle S \rangle = \{ \sigma'(1) \ | \ \sigma' \in X, \ \ \absMap(\sigma'(1)) = S \} $$
where $\sigma'(1)$ represents the first state of trace $\sigma'$.
Let
$\nxtState{\mvni{1}}(S) = \{ S_{1}, \ldots, S_{m} \}$,
then using Definition \ref{dfn:stepTerm} we can define the step term
$$\stShort{X\langle S \rangle}{S} =
   \stepTerm{S}{X\langle S \rangle}{\setS{S_{1}}, \ldots, \setS{S_{m}}}
$$
Clearly, $\stShort{X\langle S \rangle}{S}$ must be valid by (1) above.
We can now recursively define a set of step terms closed under
step term inclusion from $\stShort{X\langle S \rangle}{S}$ as follows.

Define
$\clSetST{X\langle S \rangle} =
   \bigcup_{i \in \nat} \clSetST{X\langle S \rangle}_{i}$,
where
$\clSetST{X\langle S \rangle}_{i}$
is defined recursively:
$\clSetST{X\langle S \rangle}_{0} = \{ \stShort{X\langle S \rangle}{S} \}$
and
$$\clSetST{X\langle S \rangle}_{i+1} =
   \{ \stShort{\setS{V_{j}}}{V_{j}} \ | \
   \stepTerm{V}{\Gamma}{\setS{V_{1}}, \ldots, \setS{V_{r}}}
   \in \clSetST{X\langle S \rangle}_{i}, \ V_{j} \in \{V_{1}, \ldots, V_{r} \} \}.$$
Clearly, the set $\clSetST{X\langle S \rangle}$ is closed under step term
inclusion by construction.
Also note that it can only contain valid step terms;
this follows from (1) above and the fact that if
$\stShort{\Gamma}{S}$
is a valid step term then any new step term
$\stShort{\Gamma \cup \{ S' \} }{S}$
formed by adding an additional state $S' \in \setStAbs{S}$
must also be valid.
It therefore follows that
for each $S \in \ssEntity{\mvni{1}}$ we know that each
step term $\stShort{\Gamma}{S_{i}} \in \clSetST{X\langle S \rangle}$
must occur in the initial family $C$ of sets of step terms
used in the decision procedure, i.e.
$\stShort{\Gamma}{S} \in C(S)$.
Since none of these step terms can be removed from $C$ by the closure property
it follows that the decision procedure
{\tt checkAsynAbs($MV_{1}$,$MV_{2}$,$\absMap$)}
must terminate returning $true$.
\pfbox
\\

\section{Case Study: The Regulation of Tryptophan Biosynthesis}
\label{case} In this section we present a detailed case study which
illustrates the abstraction techniques developed in the previous
sections. Our case study is based on identifying abstractions for a
published MVN model of the regulatory system used to control the
biosynthesis of tryptophan in {\it E. coli} \cite{Simao2005}.
Tryptophan is an amino acid which is essential for the development
of {\it E. coli}. However, the synthesis of tryptophan is resource
intensive and for this reason is carefully controlled to ensure it
is only synthesised when no external source is available. 
The regulatory network that controls the biosynthesis of tryptophan by
{\it E. coli} has been extensively studied (see for example \cite{SenLiu1990,Santillan2001}).
\begin{figure}[h]
\centering
\begin{tabular}{c@{\qquad}c}
  \includegraphics[width=0.3\textwidth,keepaspectratio]{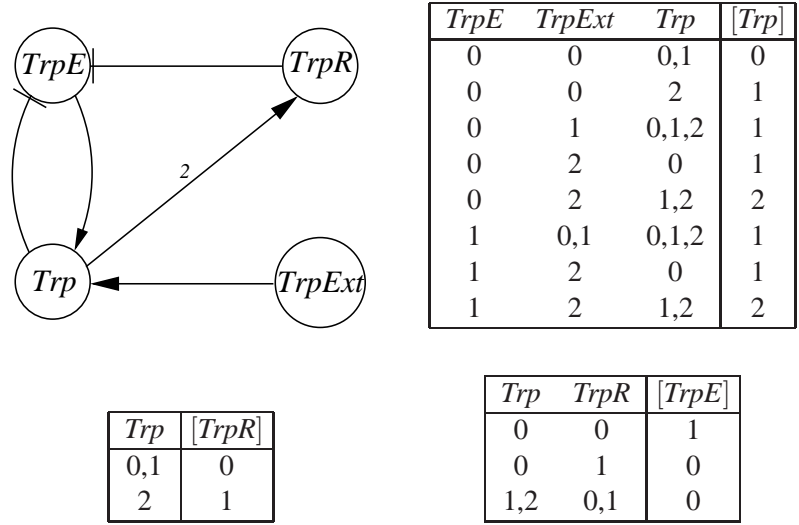}

  &
\begin{tabular}[b]{|ccc|c|}
        \hline $\TrpE$  & $\TrpExt$ & $\Trp$ & $\!\!\nextSt{\Trp}\!\!$\\
        \hline
        0   & 0 & 0,1      & 0 \\
        0   & 0 & 2        & 1 \\
        0   & 1 & 0,1,2    & 1 \\
        0   & 2 & 0        & 1 \\
        0   & 2 & 1,2      & 2 \\
        1   & 0,1 & 0,1,2  & 1 \\
        1   & 2   & 0      & 1 \\
        1   & 2   & 1,2    & 2 \\
        \hline
\end{tabular}
\\
\\
\begin{tabular}[b]{|c|c|}
        \hline $\Trp$  &  $\!\!\nextSt{\TrpR}\!\!$\\
        \hline
        0,1     & 0 \\
        2       & 1 \\
        \hline
\end{tabular}
&

\begin{tabular}[b]{|cc|c|}
        \hline $\Trp$  & $\TrpR$ & $\!\!\nextSt{\TrpE}\!\!$\\
        \hline
        0   & 0     & 1 \\
        0   & 1     & 0 \\
        1,2 & 0,1   & 0 \\
        \hline
\end{tabular}

\end{tabular}
\caption{
An MVN model $\trpMV$ of the regulatory mechanism for the biosynthesis of tryptophan in {\it E. coli}
(from \cite{Simao2005}).
The state transition table for $\TrpExt$ has been
omitted as this is a simple input entity.
Note that the state transition tables use a
shorthand notation where an entity is allowed to
be in any of the states listed for it in a particular row.}
\label{fig:exTrp}
\end{figure}
%

Consider the MVN model $\trpMV$ for tryptophan biosynthesis presented in Figure \ref{fig:exTrp}
which is taken from \cite{Simao2005}.
It consists of four regulatory entities:
$\TrpE$ -- a Boolean input entity indicating the presence of the activated enzyme required for synthesising tryptophan;
$\TrpR$ -- a Boolean entity indicating if the repressor gene for
tryptophan production is active;
$\TrpExt$ -- a ternary entity indicating the level of tryptophan in the external medium; and
$\Trp$ -- a ternary entity indicating the level of tryptophan within the bacteria.
Note the above entity order is used when displaying global states for $\trpMV$.
We can see from the model that the presence of tryptophan in the external medium $\TrpExt$ directly affects the level of tryptophan within the bacteria $\Trp$
and that the activated enzyme $\TrpExt$ is required to synthesise tryptophan.
The presence of tryptophan within the bacteria deactivates the enzyme $\TrpE$ and at higher-levels
also activates the repressor $\TrpR$ which then acts to inhibit the production of the enzyme $\TrpE$.

The state space for the $\trpMV$ consists of $36$ global states and for this reason
we do not reproduce its state graph here.
Instead we simply note that the asynchronous state graph for $\trpMV$
comprises three disjoint graphs based on the following three attractors:
$0000 \upStepAsy 1000 \upStepAsy 1001 \upStepAsy 0001 \upStepAsy 0000$;
$0011$; and
$0122$.
To identify abstractions for $\trpMV$ we begin by defining appropriate state mappings for the non-Boolean entities $\TrpExt$ and $\Trp$ as follows:
$$
\mapping{\Trp} = \{0 \mapsto 0, 1 \mapsto 1, 2 \mapsto 1\}, \ \ \ \
\mapping{\TrpExt} = \{0 \mapsto 0, 1 \mapsto 1, 2 \mapsto 1\}.
$$
These can then be combined into an abstraction mapping
$$\absMap = \langle I_{\TrpE}, I_{\TrpR}, \mapping{\TrpExt}, \mapping{\Trp} \rangle.$$

Following the approach presented in \cite{banksSteggles2010}, we first apply this abstraction
mapping to $\trpMV$ to produce a set $\absMap(\trpMV)$ of candidate abstraction models.
By analysing $\absMap(\trpMV)$ we are able to establish that there are
8 possible candidate abstraction models (we have 4 choices for next-state of $\TrpR$ and 2 choices for $\Trp$).
After investigating these candidate models we were able to identify
one valid asynchronous abstraction $\trpBool$ (which is presented in Figure \ref{fig:exTrpBool}) for $\trpMV$
under $\absMap$ using the decision procedure
{\tt checkAsynAbs(}$\trpBool$, $\trpMV$, $\absMap${\tt )}.
Note that since $\trSetA{\trpBool}$ and $\absMap(\trSetA{\trpMV})$ are in fact finite trace sets
in this case we were able to verify the result
$\trpBool \absA{\absMap} \trpMV$,
by checking that
$\trSetA{\trpBool} \subseteq \absMap(\trSetA{\trpMV})$ holds.
\\

\begin{figure}[h]
\centering
\begin{tabular}{c}

\begin{tabular}{|c|c|}
        \hline $\Trp$  & $\!\!\nextSt{\TrpR}\!\!$\\
        \hline
        0,1   & 0 \\
        \hline
\end{tabular}
\\ \\
\begin{tabular}{c@{\qquad}c}

\begin{tabular}{|ccc|c|}
        \hline $\TrpE$  & $\TrpExt$ & $\Trp$ & $\!\!\nextSt{\Trp}\!\!$\\
        \hline
        0   & 0 & 0,1      & 0 \\
        0   & 1 & 0,1      & 1 \\
        1   & 0,1 & 0,1    & 1 \\
        \hline
\end{tabular}

&

\begin{tabular}{|cc|c|}
        \hline $\Trp$  & $\TrpR$ & $\!\!\nextSt{\TrpE}\!\!$\\
        \hline
        0   & 0     & 1 \\
        0   & 1     & 0 \\
        1   & 0,1   & 0 \\
        \hline
\end{tabular}
\\
\end{tabular}

\end{tabular}
\caption{
The asynchronous abstraction $\trpBool$ identified for $\trpMV$ under the state mappings
$\mapping{\Trp} = \{0 \mapsto 0, 1 \mapsto 1, 2 \mapsto 1\}$ and
$\mapping{\TrpExt} = \{0 \mapsto 0, 1 \mapsto 1, 2 \mapsto 1\}$.}
\label{fig:exTrpBool}
\end{figure}
%
The state graph for $\trpBool$ consists of two disjoint graphs and has two attractors:
$0000 \upStepAsy 1000 \upStepAsy 1001 \upStepAsy 0001 \upStepAsy 0000$; and
$0011$.
It therefore successfully captures two of the three attractors present in $\trpMV$.

\section{Conclusions}
\label{concl} In this paper we have developed an abstraction theory
for asynchronous MVNs based on extending the ideas developed for
synchronous MVNs \cite{banksSteggles2010} and defined what it means
for an MVN to be correctly abstracted by a simpler MVN with the same
network structure but smaller state space. The abstraction approach
used is based on an {\it under--approximation} approach
\cite{Clarke1994,Pelanek2006} in which an abstraction captures a
subset of the behaviour of the original MVN. We showed that this
approach allows positive reachability properties of an  MVN to be
inferred from a corresponding asynchronous abstraction and that all
attractors of an asynchronous abstraction correspond to attractors
in the original MVN. An alternative approach would be to use an {\it
over--approximation} approach \cite{Clarke1994,Pelanek2006,Clarke03}
in which false positives can arise. However, the construction of an
abstraction model which over--approximates an MVN's behaviour
appears to be problematic if we wish to remain within the MVN
framework (see Section \ref{sec:asynAbs} for a discussion of this).

Directly checking asynchronous abstractions turned out to be
problematic given that an asynchronous MVN may have an infinite set
of traces which makes it infeasible to directly check trace
inclusion. To address this we developed a decision procedure for
checking asynchronous abstractions based on the finite state graph
of an asynchronous MVN. The decision procedure used {\it step terms}
to denote possible ways to use sets of concrete states to represent
abstract states and worked by iteratively pruning the set of step
terms until either a consistent abstract representation has been
found or the set of remaining step terms is too small to make it
feasible to continue. Importantly, we provided a detailed proof that
showed the decision procedure worked correctly.
Note that as it stands, the decision procedure is inefficient;
work is on going to refine this procedure and to use it as a basis of a tool for
abstraction checking.
Such a tool will provide the support needed to carry out more complex case studies,
for example supporting the work currently underway to investigate abstractions for the
relatively complex MVN model of the carbon starvation response in {\it E. coli} presented in \cite{banksSteggles2007}.

We illustrated the abstraction theory and techniques developed by considering a
detailed case study based on identifying a Boolean abstraction for an asynchronous MVN model of
the regulatory system used to control the biosynthesis of
tryptophan in {\it E. coli}.
The abstraction found proved to faithfully represent the behaviour of the original MVN and
in particular, captured two of the three attractors known to exist in the original MVN.
The case study illustrates the potential for the abstraction theory presented and in particular, how it
allows the balance between the level of abstraction used and the tractability of analysis to be explored.

An alternative approach for abstracting MVNs is to
reducie the number of regulatory
entities in an MVN while ensuring the preservation of key properties (see \cite{Naldi09,Velizcuba2009,Naldi2011}).
This approach seems to be complimentary to the one developed here and we are currently
investigating combining these ideas.
Another possible abstraction approach would be to make use of results on modelling MVNs using Petri nets \cite{Comet05,banksSteggles2007,banksetal2010,Chaouiya2011}
and to then apply Petri net abstraction techniques (see for example \cite{Suzuki1983,Kungas2005,Wimmel2011}).
Such an approach appears promising from an analysis point of view but problematic in that the
resulting Petri net abstraction may not be interpretable as an MVN and so force the modeller
to explicitly use a different modelling formalism.
%

One interesting area for future work is to investigate automatically
constructing abstractions for a given MVN and abstraction mapping.
Some initial work on restricting the search space for such
abstractions can be found in \cite{banksSteggles2010} but more work
is needed here. One idea is to consider developing refinement
techniques similar to those of {\it CEGAR (Counterexample Guided
Abstraction Refinement) } \cite{Clarke03} and other abstraction
refinement techniques \cite{Pelanek2006}.
Closely linked to this idea is the notion of a maximal abstraction, that is an abstraction which captures the largest possible behaviour of the original MVN with respect to all other possible abstractions for the given abstraction mapping.
In future work we intend to investigate developing such a notion and in particular, consider how to automate the construction of
such maximal abstractions.

\medskip
\noindent\textbf{Acknowledgments.} We would like
to thank Richard Banks and Maciej Koutny for their advice
and support during the preparation of this paper.
We would also like to thank the anonymous referees for their
very helpful comments and suggestions.

\bibliographystyle{mecbic}

\end{document}